\documentclass[aps,twocolumn,amsmath,amssymb]{revtex4}
\usepackage{amssymb}
\usepackage[dvips]{graphicx}
\usepackage{dcolumn}
\usepackage{bm}
\usepackage{epsfig}
\usepackage[latin1]{inputenc}

\begin{document}

\title{Anharmonic transitions in nearly dry L-cysteine I}

\author{T. A. Lima$^{1}$}

\author{E. T. Sato$^{1}$}

\author{E. T. Martins$^{1}$}

\author{P. Homem-de-Mello$^{1}$}

\author{A. F. Lago$^{1}$}

\author{M. D. Coutinho-Neto$^{1}$}

\author{F. F. Ferreira$^{1}$}

\author{C. Giles$^{2}$}

\author{M. O. C. Pires$^{1}$}

\author{H. Martinho$^{1}$}

\affiliation{$^{1}$Centro de Ciências Naturais e Humanas, UFABC, Rua Santa Adélia 166,09210-170, Santo André, São Paulo, Brazil}

\affiliation{$^{2}$Instituto de Física ''Gleb Wataghin'', UNICAMP, 13083-970,
Campinas, São Paulo, Brazil}

\begin{abstract}

Two special dynamical transitions of universal character have been recently observed in macromolecules at $T_{D}\sim 180 - 220$ K and $T^{*}\sim 100$ K. Despite their relevance, a complete understanding of the nature of these transitions and their consequences for the bio-activity of the macromolecule is still lacking. Our results and analysis concerning the temperature dependence of structural, vibrational and thermodynamical properties of the orthorhombic polymorph of the amino acid L-cysteine (at a hydration level of $3.5\%$)  indicated that the two referred temperatures define the triggering of very simple and specific events that govern all the biochemical interactions of the biomolecule: activation of rigid rotors ($T<T^{*}$ ), phonon-phonon interactions with phonons of water dimer ($T^{*}<T<T_{D}$), and water rotational barriers surpassing ($T>T_{D}$).

\end{abstract}

\maketitle

Recently two special dynamical transitions, apparently of universal character, were observed in macromolecules as lysozyme, myoglobin, bacteriorhopsin, DNA, and RNA (see, e.g., refs.\cite{3,5}). With cooling, the first transition occurred at $T_{D}\thickapprox 180-220$ K at hydration level $h>18\%$. It is related to a transition from anharmonic to harmonic regime\cite{5}. The most remarkable character of this transition is its direct correlation to the biological activity\cite{3,5}. On the basis of quasi-elastic neutron scattering (QENS) measurements on lysozyme, Chen et al.\cite{7} interpreted this transition as a fragile-to-strong dynamic crossover where the structured water makes a transition from a high-density to a low-density state. This scenario was based on studies that advocated a second critical point of water (liquid-liquid critical point) in confined supercooled water\cite{8}. However, a recent high resolution QENS experiment performed by Dosteret al.\cite{9} in fully deuterated C-phycocyanin protein, showed no evidence of such fragile-to-strong character in the dynamical transition at $T_{D}$. As pointed out by these authors, their experimental findings are consistent with a glass transition of the hydration shell. However, the specific nature of the glassy state remains unclear. Moreover, the specific role of protein and water is still an issue. As an example, Frauenfelder et al.\cite{3} proposed that the protein dynamics is slaved to the bulk solvent fluctuations. On the other hand, the dynamics of tRNA compared to other studies on DNA and proteins by Khodadadiet al.\cite{11} contradicts the slave picture. Their results indicated the mutual influence of macromolecule and hydration water must be considered.
\begin{figure}[h!]
\includegraphics[width=6.0cm]{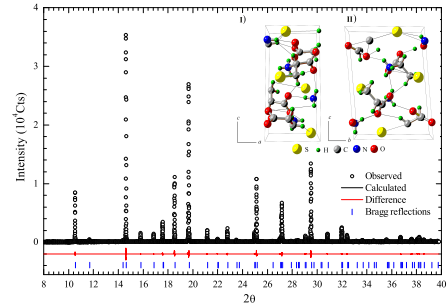}
\caption{Experimental (open circles) and calculated (black solid line) diffraction pattern of orthorhombic L-cysteine I structure.The vertical bars and red solid line at the bottom represent the Bragg reflections and the difference between observed and calculated data, respectively. The insert shows the unit cell of L-cysteine I observed along the ac (panel I) and bc (panel II) planes.}\label{fig1}
\end{figure}
The other important transition observed in macromolecules is a hydration independent sub-regime of harmonicity onset at $T^{*}\sim 100$ K\cite{5}.   The microscopic nature of this transition is still an object of debate. The discussed possibilities include quantum effects near a zero point vibration\cite{12}, and methyl group rotation associated with anharmonicity\cite{5,13}. Vibrational modes play a very important role in the conformational arrangement of different functional states and also in the energetic balance of the chemical bonds\cite{14}. Theoretical studies on the anharmonic effects on phonons show that their frequency and linewidth analyzed by the inelastic light scattering (Raman effect) could vary due to the anharmonic potential shape of the simple oscillator and also due to phonon-phonon interactions\cite{15,16}.

The present study is focused on the temperature dependence of structural, Raman-active vibrational modes and thermodynamical properties of the orthorhombic polymorph of the amino acid L-cysteine, $C_{3}H_{7}NO_{2}S$, $^{+}NH_{3}-CH(CH_{2}SH)-COO^{-}$ . This amino acid possesses a very simple chemical structure and high biological relevance. The thiol or sulfhydryl group in the residues of L-cysteine is the most chemically reactive site in proteins under physiological conditions\cite{19}. Therefore, the knowledge of its dynamics is very important.

\begin{figure}[h!]
\includegraphics[width=6.0cm]{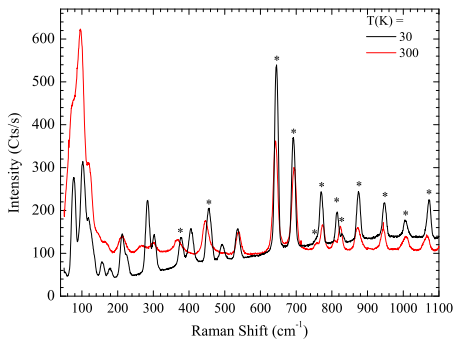}
\caption{Raman spectra at 30 K (black line) and 300 K (red line). Those bands marked with asterisk ($^{*}$) had their detailed temperature behavior.}\label{fig2}
\end{figure}

Figure 1 shows the X-ray powder diffraction pattern of the studied sample. The Rietveld refinement indicated that the predominant phase is orthorhombic ($P2_{1}2_{1}2_{1}$ space group) with unit cell parameters $a = 8.09758(8)$, $b = 12.1495(1)$, and $c = 5.4142(1)$ ${\AA}$. The unit cell (insert of Fig. 1) has 4 molecules ($z=4$ ) of L-cysteine. Our findings are consistent with others previously reported in the literature\cite{21,22}.

Figure 2 shows the Raman spectra of L-cysteine I at 30 and 300 K. A clear energy threshold at 500 cm$^{-1}$(60 meV) could be observed. All bands above this energy kept their shape with merely a softening due to temperature variation. The exception was the bands at 815 and 830 cm$^{-1}$ where an intensity transfer was observed between them. A completely diverse situation occurred below 500 cm$^{-1}$ where some bands broadened at high temperatures. \textit{Ab initio} vibrational analysis calculations were performed to interpret the Raman spectra. Since the relevance of the intermolecular interactions (cys-cys, H-bond, etc) is an important issue,  two methodologies based on the Density Functional Theory (DFT)\cite{23} were compared: free molecule calculations performed using the B3PW91 hybrid functional\cite{25} as implemented on the on Gaussian 03 software suite\cite{27}  and unit cell with periodic boundary condition calculations using a fixed unit cell volume and the BLYP functional\cite{28} augmented with corrections for the proper description of dispersion interactions\cite{30} as implemented in the CPMD code\cite{32}. Only those values in accordance with the experimental finding within $5\%$ variation were considered for the assignment. It is important to stress that the free molecule calculated frequencies that meet this criterion were only those modes related to carbon backbone torsion. Similar results were obtained by Pawlucojk et al.\cite{22}. The corresponding bands in the crystal calculation showed better agreement with experimental values with a larger set of concordant bands. As a matter of fact, most crystal modes presented contributions of almost all atoms in the unit cell with their corresponding eigenvectors showing a more complex dynamics than the free molecule case. This fact is a manifestation of the role of the intra and inter-molecular interactions in the vibrational modes description, both responsible for the strong anharmonicity of this system.

\begin{figure}[h!]
\includegraphics[width=6.0cm]{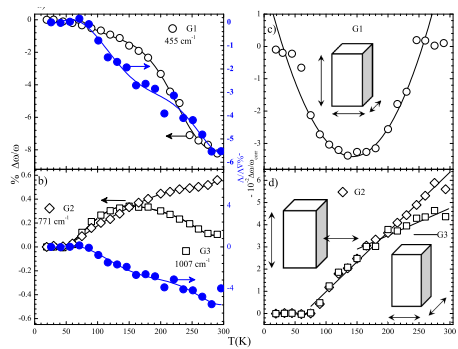}
\caption{$\%_{\Delta}\omega/\omega$ (black scale) scaled to ($\%_{\Delta}V/V$ ) (blue scale) as a function of temperature for G1 (open circles,a)), G2 ( open diamond, b)), and G3 (open square, b)) phonons. ($\%_{\Delta}\omega/\omega_{corr}$) is shown in the right side (c) and d)). The solid lines show the fitting to the corresponding polynomial functions (quadratic, c); and linear, d)). The inserts represent the unit cell motion evolved in each phonon group.}\label{fig3}
\end{figure}

We considered a detailed temperature dependence study for those modes where hydrogen bonds participate (signed with asterisks on Fig. 2). These particular vibrations are candidates to mediate the water-molecule interactions. At this point it is important to distinguish the two kinds of anharmonic contributions to the temperature dependence of the phonon frequency. The first one arises from anharmonic deviations of the single harmonic oscillator potential shape. The second one is related to phonon-phonon coupling\cite{16}. The isothermal mode Grüneisen parameter  is a parameter that quantifies the phonon anharmonicity of the first type. A good estimation for  could be obtained scaling the phonon frequency ($\Delta\omega/\omega$ ) to unit cell ($\Delta V/V$) temperature variation.

This scaling was performed on the phonons indicated by asterisks. We were able to distinguish three kinds of characteristic behaviors. Figure 3a) and b) display the representative behavior for each member of these groups. The right and left scales show the phonon frequency variation and the negative of the unit cell volume variations, respectively. The first group (G1) comprises the 455, 874, and 945 cm$^{-1}$ phonon modes. Figure 3a) (open circles) shows the representative temperature dependence for the 455 cm$^{-1}$ vibration. With heating, an overall softening for all members of this group following the lattice expansion was observed. However, the softening rate displayed a smooth increase between $\sim100$ and $250$ K compared to  $\Delta V/V$.The second group (G2) comprised the 538, 755, and 771 cm$^{-1}$ phonons. The observed behavior was the opposite of the one observed following lattice expansion, as can be seen in Fig. 3b) (open diamond) for the 771 cm$^{-1}$ mode. The third group (G3) has a single member (1007 cm$^{-1}$ phonon) and presented a very peculiar behavior (open square symbol in Fig. 3b). Up to 70 K, its frequency was almost constant. Thereafter, a sudden hardening was observed between 70 and 150 K followed by a softening. In order to carry out a detailed analysis of this contribution we subtracted the scaled  behavior from the phonon temperature dependence. The data is shown in Fig. 3c) and d). Those modes of G1 group presented a characteristic quadratic behavior (Fig. 3c). The data for G2 and G3 group phonons displayed a clear linear dependence.  Above $\sim200$ K a tendency to change was observed for G3, which presented smaller slope than the G2 one. It is important to notice that recent reports on literature claimed for the occurrence of signatures of subtle phase transitions at $\sim70$ K in L-cysteine by specific heat, Raman spectroscopy, X-ray diffraction, and neutron scattering\cite{33,34,35}.This kind of behavior can be understood by considering phonon-phonon interactions.  Balkanski, Wallis, and Haro\cite{15}  developed a detailed theory of the higher order phonon anharmonic decay in order to explain the temperature dependence of the frequency and linewidth of the optical phonons probed by Raman scattering. Since the Debye temperature ($\Theta_{D}$) of L-cysteine is very low ($<5$ K as estimated by our specific heat data in Fig. 4), we could use the high temperature limit of the Balkanski theory.  Using this approximation, the decay of one phonon with frequency  in two and three phonons will provide specific temperature contributions given by

 $\Delta\omega/\omega\varpropto T$, for a three phonon decay process, and

 $\Delta\omega/\omega\varpropto T^{2}$, for a four phonons decay process.

Thus, in order to explain the observed temperature behavior of the three afore-mentioned groups one needs to find phonons that fit the sum rule and would participate in the decaying process. In the present context the water vibrational bands are natural candidates. In fact, our vibrational analysis calculations of water dimer showed that water dimer presented a set of low frequency  intermolecular vibrations at 93 cm$^{-1}$ (donor twisting/acceptor rocking); 161 cm$^{-1}$ (donor rocking/acceptor wagging); 195 cm$^{-1}$ (acceptor rocking); and  251 cm$^{-1}$ (O-O stretching)which could participate in the decay process. For G1 phonons, we can establish the following correlation

 $455\rightarrow 93+161+196$ cm$^{-1}$

 $874\rightarrow 455+251+161$ cm$^{-1}$

 $945\rightarrow 455+2\times251$ cm$^{-1}$, which is correct within $2\%$ precision. Each phonon decays into a combination of water and/or G1-member phonon. Likewise, the correlation

 $538\rightarrow 455+93$ cm$^{-1}$

 $755\rightarrow 538+195$ cm$^{-1}$

 $771\rightarrow 538+251$ cm$^{-1}$, which is correct within $3\%$ precision could be established for G2 phonons. Finally, for G3 ($1007$ cm$^{-1}$), we have $1007\rightarrow 755+251$ cm$^{-1}$, correct within $2\%$ precision.

The eigenvectors of the vibrational modes for each group also presented characteristic patterns. The vibrational modes of G1 group were characterized by a stretching motion of the unit cell along the three crystallographic axes (see inset of Fig. 3c). The atomic motions in the G2 and G3 groups resulted in the stretching of the cell along two perpendicular axes (see inset of Fig. 3d)). For G2, the movement was along $b$ and $c$ directions. In the G3 mode the movement was confined in the $ab$ plane. This symmetry may be relevant to the understanding of the interactions between L-cysteine and water modes. In fact, our results concerning the water dimer vibrational modes indicated that the dipole derivative unit vector lied along the donor oxygen and acceptor hydrogen direction (main component along the O-O axis) for 93 and 161 cm$^{-1}$ modes, being restricted to the perpendicular directions for the other modes. Changes in the relative orientation of the hydrogen atoms of the water dimers might have a strong influence on the coupling to the L-cysteine modes. The rearrangement of the hydrogen bonding pattern due to thermally activated tunneling pathways among configurations of water dimers with exchange of acceptor and donor hydrogens\cite{36}  is an option that should be considered. Our results could be consistently explained once considering the presence of one water dimer in the L-cysteine unit cell. Since the G2 phonons do not change their behavior above 100 K (Fig. 3d) we argue that the dimers have their O-O axis along the a-direction. At 200 K, there is enough thermal energy to overcome the three water dimer tunneling pathways (bifurcation, interchange, and acceptor switching, see Keutsche al. in ref. \cite{36}. This might explain the slope change observed on the G3 phonon frequency behavior (Fig. 3d).

\begin{figure}[h!]
\includegraphics[width=5.0cm]{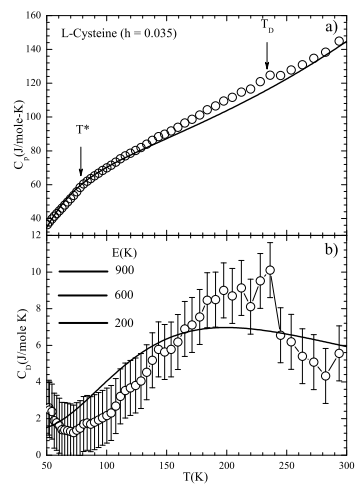}
\caption{a)$C_{p}$ showing two tiny transitions at $T^{*}\sim80$ K and $T_{D}\sim230$ K.  The solid lines are simulations as described in the text.}\label{fig4}
\end{figure}

The specific heat ($C_{p}$) results are shown in Fig.4. Fig. 4a) shows the data between $50$ and $300$ K. Very subtle transitions are seen at $220$ K and $80$ K as broad and tiny bumps near these temperatures. Calorimetric measurements on orthorhombic L-cysteine in the literature reported only a subtle transition at $T^{*} = 70$ K\cite{33}. Incoherent inelastic neutron scattering on monoclinic L-cysteine\cite{35} revealed an anharmonic transition at $T_{D} = 150$ K. Thus, it is reasonable to associate the two transitions observed for L-cysteine to the $T_{D}$ and $T^{*}$ ones, which is commonly seen in proteins and other  macromolecules.

According to the literature,  the low temperature transition at $T^{*}$ is related to methyl group rotations in proteins\cite{3}.  Since the specific heat of anisotropic rigid rotors is well known\cite{37}, this hypothesis could be easily checked. Thus, considering additional linear (electronic) and cubic (phononic) contributions, the specific heat is written as

$C_{p}=\gamma T+\beta T^{3}+C_{rotor}(T,I_{xy},I_{z})$

where $C_{rotor}(T,I_{xy},I_{z}$ is given by expressions (3) and (4) of ref.\cite{37},  with $I_{xy}$ and $I_{z}$ the inertia momenta in the $xy-$plane and $z-$axis, respectively. The simulated curve is shown as a solid line in Fig. 4a). The set of parameters that furnished the best simulated curve for $T< 150$ K were found to be $\gamma=0.205$ J/mole K$^{2}$;$\beta=1.15\times10^{-6}$ J/mole K$^{4}$; and $I_{xy}/I_{z}=0.72$ . The high $\gamma$-value is consistent and expected for an isolating material. The Debye temperature estimated from $\beta$ value was $\Theta_{D}<5$K. The ratio $I_{xy}/I_{z}$ found is consistent with a prolate symmetric top rotor (CH$_{2}$ rotation).

It is a remarkable fact the observation of the $T_{D}$ transition at very low water content ($3.5\%$) as described herein. This water concentration is equivalent to 1 water molecule per unit cell in average. This finding is inconsistent with the fragile-to-strong picture of Chen et al\cite{7} , since it is not possible to treat the aqueous content as very low density aqueous phase. In a similar way, the idea of a glassy transition of the hydration shell\cite{8,9} is not supported. Thus, another explanation concerning the origin of the dynamical transition should be considered. By subtracting the above-mentioned simulated curve from the raw data, as it can be seen in Fig. 4b), we obtained an estimate of the heat capacity evolved in the transition $T_{D}$. The broad transition resembles the shape of a Schottky anomaly due to thermal population in some specific energy levels. One possible interpretation is to consider the anomaly as originated from thermal population of tunneling splitting energy levels in configurations of water dimers to be consistent with the analysis of Raman data in Fig. 3d). The best choice was obtained using the three-level scheme with energies $E_{1}=250$K (174 cm$^{-1}$), $E_{2}=600$ K (420 cm$^{-1}$), and $E_{3}=900$ K (629 cm$^{-1}$). The agreement between the experimental data and simulation was reasonable considering the data dispersion. Keutsch et al.\cite{36} calculations (gaseous phase) indicated that the three distinct water dimer low barrier tunneling pathways acceptor switching, interchange tunneling, and bifurcation tunneling have energies of $\sim 157, 207,$ and $394$ cm$^{-1}$, respectively. The extra constraints imposed by the environmental hydrogen bonds will increase the tunneling splitting. Therefore, the energies obtained in the simulation compared to the gaseous phase dimer are expected to be greater.

This work was supported by Fapesp/Brazil, CNPq/Brazil

{}
\end{document}